\documentclass[review]{elsarticle}

\usepackage{lineno,hyperref}
\modulolinenumbers[5]
\usepackage{graphicx,color}
\usepackage{wrapfig}
\usepackage{dcolumn}
\usepackage{gensymb}

\journal{arXiv}

\begin{document}

\begin{frontmatter}

\title{Exploring the effects of ultraviolet radiation on the properties of Fe$_{3}$O$_{4}$/PANI nanostructures}

\author{Luana Hildever$^{1}$, Thiago Ferro$^{1}$, Francisco Estrada$^{2}$, and José Holanda$^{1, 3, }$\corref{mycorrespondingauthor}}
\cortext[mycorrespondingauthor]{Corresponding author: joseholanda.silvajunior@ufrpe.br}
\address{$^{1}$Programa de Pós-Graduação em Engenharia Física, Universidade Federal Rural de Pernambuco, 54518-430, Cabo de Santo Agostinho, Pernambuco, Brazil.}
\address{$^{2}$Facultad de Biología, Universidad Michoacana de San Nicolas de Hidalgo, Av. F. J. Mujica s/n Cd. Universitaria, Morelia, Michoacán, México.}
\address{$^{3}$Unidade Acadêmica do Cabo de Santo Agostinho, Universidade Federal Rural de Pernambuco, 54518-430, Cabo de Santo Agostinho, Pernambuco, Brazil.}

\begin{abstract}
The properties of any material are the basis for the most diverse applications of science, which allows for the dazzling development of new technologies. In this work, we studied the main properties of nanostructures synthesized under ultraviolet radiation. We show that the nanostructures produced are of excellent quality, which is evidenced by the high stability and quality of the magnetic properties.
\end{abstract}

\begin{keyword}
\texttt{}Ultraviolet\sep radiation\sep nanostructures\sep magnetic properties
\end{keyword}
\end{frontmatter}
\vspace{1.5cm}

\section{Introduction}

The stimulating area of magneto-optical effects has presented discoveries of new phenomena with enormous possibilities for applications [1-7]. A common point of applications are the interactions involved [6]. The fundamentals of these effects are quantum mechanics, the Pauli distribution, interactions of Heisenberg, Zeeman, Rashba, spin-orbit, chemical electronegativity, covalent bonds, double valence, and the definition of chemical potential, among many others [8-10]. In this context, three iron oxides stand out: magnetite, maghemite, and hematite. 
\begin{itemize}
	\item
	Magnetite (Fe$_{3}$O$_{4}$) has a face-centered cubic structure with a lattice parameter of \textit{a} = 0.839 nm and a saturation magnetization on the order from 90 to 100 emu/g [11, 12]. Magnetite differs from other structures of iron oxides in that it contains divalent iron ions Fe$^{2+}$ and trivalent iron ions Fe$^{3+}$. Magnetite has an inverse spinel crystal structure, where its tetrahedral sites have ions Fe$^{3+}$, while the octahedral site has both ions Fe$^{3+}$ (S = 5/2) and Fe$^{2+}$ (S = 2) [11-16]. The arrays of the octahedron and tetrahedron, as well as the ordering of the spins in the magnetite unit cell present that the spin of the eight ions Fe$^{3+}$ in the sites cancel with the eight ions Fe$^{3+}$ in the sites. Therefore, the resulting magnetic moment is due exclusively to the Fe$^{2+}$ ions, which have spin S = 2. Furthermore, magnetite has a Curie temperature of 850 K. 
	\item
	Maghemite ($\gamma$Fe$_{2}$O$_{3}$) has a similar structure to magnetite, such that it has an inverse spinel structure forming a face-centered cubic lattice with a lattice parameter of \textit{a} = 0.834 nm and a saturation magnetization between 70 and 85 emu/g [11, 12]. Maghemite is the equivalent of oxidized magnetite, where during the process of oxidation, an ion Fe$^{2+}$ leaves of the site, leaving a vacancy in the crystalline lattice, and another ion Fe$^{2+}$ transforms into ion Fe$^{3+}$; this is the main difference between maghemite and magnetite, that is, iron is in the trivalent state [17-19]. The formula for maghemite is $\gamma$Fe$_{2}$O$_{3}$ but may be re-written by multiplying it by 4/3, which results in Fe$_{8/3}$O$_{4}$, which emphasizes the similarity between the structure of maghemite and of magnetite [13-16]. Regarding temperature, maghemite is ferrimagnetic at room temperature and transforms into hematite ($\alpha$Fe$_{2}$O$_{3}$) at temperatures above 800 K.
	\item
	Hematite ($\alpha$Fe$_{2}$O$_{3}$) has a hexagonal unit cell with lattice parameters \textit{a} = 0.503 nm and \textit{c} = 1.375 nm and has a saturation magnetization of the order from 0.2 to 0.4 emu/g [13, 14]. Furthermore, at room temperature, it becomes weakly ferromagnetic, at 260 K (Morin temperature), undergoes a phase transition to an antiferromagnetic state, and above 956 K is paramagnetic [13, 14].
\end{itemize}
	
The structure of polymers consists of the repetition of small monomeric units connected through covalent bonds, resulting in large molecular chains, whose properties depend on the nature of the monomeric units, chain size, and crystallinity, among other properties [20, 21]. Polyaniline (PANI) has received attention due to its ease of synthesis, low cost, chemical stability, and high conductivity compared to other conductive polymers [20, 21]. For the preparation of PANI, the most used methods are chemical and electrochemical, the latter consisting of the oxidation of the aniline monomer (C$_{6}$H$_{5}$NH$_{2}$) in acidic electrolyte solutions such as H$_{2}$SO$_{4}$, HCl, and HNO$_{3}$, among many others. Regardless of the synthesis process, the structure of polyaniline in base form (undoped) has three fundamental states: oxidized, reduced, and oxidized-reduced. A completely oxidized state is called pernigranilin; a completely reduced state is called leucoeseraldine; and an oxidized-reduced state with 50\% oxidation and 50\% emeraldine base is called an intermediate state [22]. These three states are all insulating. However, to form the emeraldine salt, which is the conducting form of PANI, the emeraldine base reacts with the acid, producing protonation of the nitrogen atoms that bind to the benzenoid and quinoid rings, which separate to form polarons that will rise to as expected [20, 21]. The first PANI structure containing Fe$_{3}$O$_{4}$ nanoparticles was synthesized by mixing an aqueous Fe$_{3}$O$_{4}$ solution with an emeraldine base [22], demonstrating that the superparamagnetic properties observed in the structure are due to the Fe$_{3}$O$_{4}$ nanoparticles incorporated into PANI. Furthermore, we found that the properties of these systems (polyaniline and iron oxides) depend directly on particle size, interactions, and temperature. Due to this, several research groups began to synthesize these composites by modifying the synthesis routes to study the magnetic and conductive characteristics [20, 21, 22].

In recent years, structures containing iron oxide and PANI nanostructures have been extensively studied, due to the ideal combination between structural properties and polymeric conductivity [20, 21, 22]. Recent studies show that the Fe$_{3}$O$_{4}$/PANI structure has interesting electrical properties that can be used in applications such as data storage, ferrofluids, optoelectronics, spintronics, biomedical applications such as drug administration, degradation of dyes in the textile industry, among others. [12, 13, 14, 20, 21, 22]. In this work, we studied the main properties of Fe$_{3}$O$_{4}$ nanostructures, highlighting the quality of the samples, mainly in terms of magnetic properties.

\section{Results}

\subsection{Synthesis}

For the synthesis of Fe$_{3}$O$_{4}$/PANI nanostructures, we used commercial spherical iron oxide nanoparticles from Sigma Aldrich and the following procedure: (i) 0.1 g of nanoparticles were placed in a test tube and mixed with 6 mL of aniline sulfate solution (0.5 mol/L aniline and 1 mol/L H$_{2}$SO$_{4}$). (ii) The resulting solution was placed in a QUIMIS model Q261M magnetic stirrer with temperature heater at 1500 rpm for 3.5 hours, at the same time under ultraviolet (UV) light from a Cole-Parmer 97620-42 lamp, which emits radiation pure UV with a wavelength of 365 nm and a power of 400 W. We used a dimmer to control the intensity of the UV radiation from the lamp by setting the power at 360 W. We fixed the temperatures from the magnetic stirrer at 310 K, 330 K, and 350 K. We carried out the process in a closed environment, where we used two temperature controllers, one from model 9700 from Scientific Instruments and the other with a differential thermocouple to guarantee precise temperature control. A sample was taken from the preparation environment every 30 minutes during the synthesis. (iii) The nanostructures were centrifuged and washed several times with distilled water and acetonitrile until the supernatant solution became transparent. Then, the nanostructures were dried in an oven at temperatures from 310 K, 330 K, and 350 K as per item ii). We controlled the entire process through a graphical interface using a Python program. \textbf{Fig. 1} shows the experimental setup used in the synthesis of our samples.
\begin{figure}[h]
	\vspace{0.1mm} \hspace{0.1mm}
	\begin{center}
		\includegraphics[scale=0.24]{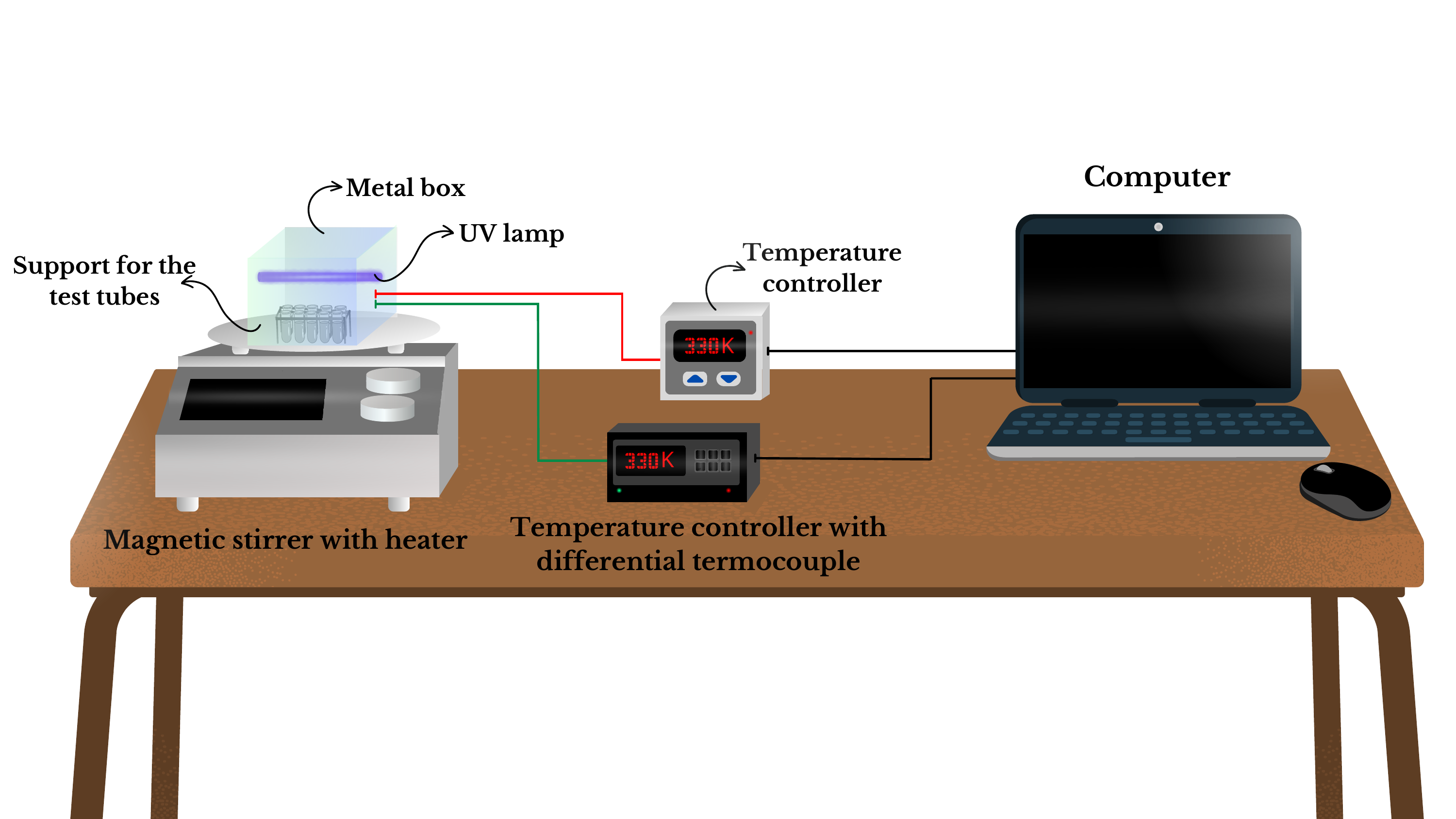}
		\caption{\label{arttype}(Color online) Experimental setup for the synthesis of Fe$_{3}$O$_{4}$/PANI nanostructures. We highlight the monitoring carried out with two temperature controllers.}
		\label{puga}
	\end{center}
\end{figure}

\subsection{Structural characterization}

In this topic, we will discuss the structural characterization of our samples through X-ray diffraction measurements using a Shimadzu XRD-7000 diffractometer with a radiation source K$\alpha_{Cu}$ ($\lambda$ = 0.1542 nm), voltage of 40 kV and current of 30 mA. The characterized samples are Fe$_{3}$O$_{4}$/PANI nanostructures treated at 310 K, 330 K, and 350 K under UV radiation for different times from synthesis t. For each temperature, there are seven samples whose reaction time varied between t = 0 min. to t = 180 min. in 30 minutes time intervals. \textbf{Fig. 2 (a)} shows the X-ray diffractograms for pure PANI and magnetite (Fe$_{3}$O$_{4}$). According to the results shown in \textbf{Figs. 2 (b)}, \textbf{(c)}, and \textbf{(d)}, the Fe$_{3}$O$_{4}$/PANI nanostructures are crystalline.
\begin{figure}[h]
	\vspace{0.1mm} \hspace{0.1mm}
	\begin{center}
		\includegraphics[scale=0.38]{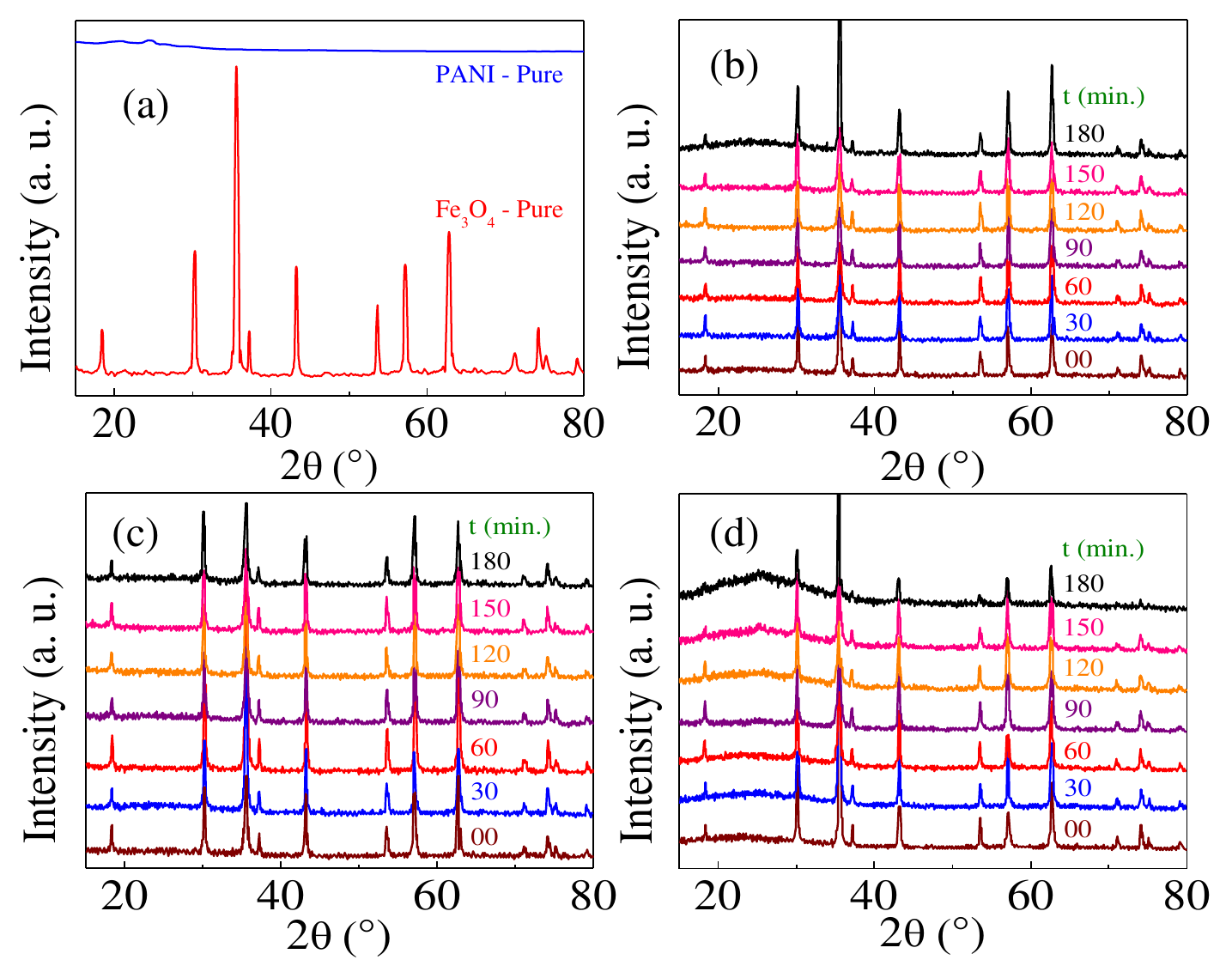}
		\caption{\label{arttype}(Color online)(a) X-ray diffraction for pure Fe$_{3}$O$_{4}$ and PANI samples. (b), (c) and (d) X-ray diffraction for Fe$_{3}$O$_{4}$/PANI samples synthesized under UV radiation with a wavelength of 365 nm. It also shows the different times t and temperatures used 310 K (b), 330 K (c), and 350 K (d) during the synthesis.}
		\label{gata}
	\end{center}
\end{figure}

\subsection{Diameter}

The average crystallite diameters were calculated from the diffractograms using the Sherrer equation 1, defined by equation (a),
\begin{equation} 
	d_{DRX} = \frac{k\lambda}{\beta cos\theta},
	\label{1}
\end{equation}
where $d_{DRX}$ is the average diameter of the crystallite, and $k$ is a constant that depends on the shape of the particle. Thus, $k$ = 1 (spherical particle), $\lambda$ is the wavelength of the incident radiation (K$\alpha_{Cu}$ = 0.1542 nm), and $\beta$ is the width at half height of the most intense diffracted peak located at 2$\theta$ [1]. $\beta$ was obtained by fitting the highest intensity peak with a Pseudo-Voigt function, which represents a combination of a Gaussian and a Lorentzian function. \textbf{Table 1} shows that the average diameter for samples synthesized all the temperatures.
\begin{table}[h]
	\caption{\label{tab:table4}{Average crystallite size estimated through X-ray diffraction for Fe$_{3}$O$_{4}$/PANI nanostructures with different times from synthesis under UV radiation with a wavelength of 365 nm and synthesis temperatures of 310 K, 330 K, and 350 K.}}
	\begin{center}
		\begin{tabular}{c|ccc}
			Temperature & 310 K & 330 K & 350 K \\
			\hline
			\hline
			Time of synthesis (min.) &  & Diameter (nm) & \\
			\hline
			0 & 44 $\pm$ 1.2 & 44 $\pm$ 1.2 & 44 $\pm$ 1.2 \\
			30 & 44 $\pm$ 1.0 & 44 $\pm$ 1.2 & 44 $\pm$ 1.1  \\
			60 & 43 $\pm$ 1.6 & 44 $\pm$ 1.2 & 43 $\pm$ 1.3 \\
			90 & 42 $\pm$ 1.4 & 44 $\pm$ 1.1 & 42 $\pm$ 1.2 \\
			120 & 41 $\pm$ 1.7 & 43 $\pm$ 1.4 & 41 $\pm$ 1.6 \\
			150 & 40 $\pm$ 1.1 & 43 $\pm$ 1.3 & 40 $\pm$ 1.4 \\
			180 & 39 $\pm$ 1.3 & 42 $\pm$ 1.6 & 39 $\pm$ 1.7 \\
		\end{tabular} 
	\end{center}
\end{table}

In \textbf{Fig. 3(a)}, we show a micrograph of a crystalline Fe$_{3}$O$_{4}$ nanoparticle obtained using a Transmission Electron Microscope (TEM) model 300 kV FEG Tecnai. On the other hand, in \textbf{Fig. 3(b)}, we present a micrograph of the Fe$_{3}$O$_{4}$/PANI nanostructures obtained using a Scanning Electron Microscopy (SEM) model 200 kV FEG Quanta. We observed that the Fe$_{3}$O$_{4}$/PANI nanostructures obtained via SEM are agglomerates of nanoparticles that demonstrate the insertion of Fe$_{3}$O$_{4}$ nanoparticles into PANI. The Fe$_{3}$O$_{4}$/PANI nanoparticles in \textbf{Fig. 3(b)} were synthesized under UV radiation with a wavelength of 365 nm at a temperature of 330 K and a time of synthesis of t = 60 minutes.
\begin{figure}[h]
	\vspace{0.1mm} \hspace{0.1mm}
	\begin{center}
		\includegraphics[scale=0.38]{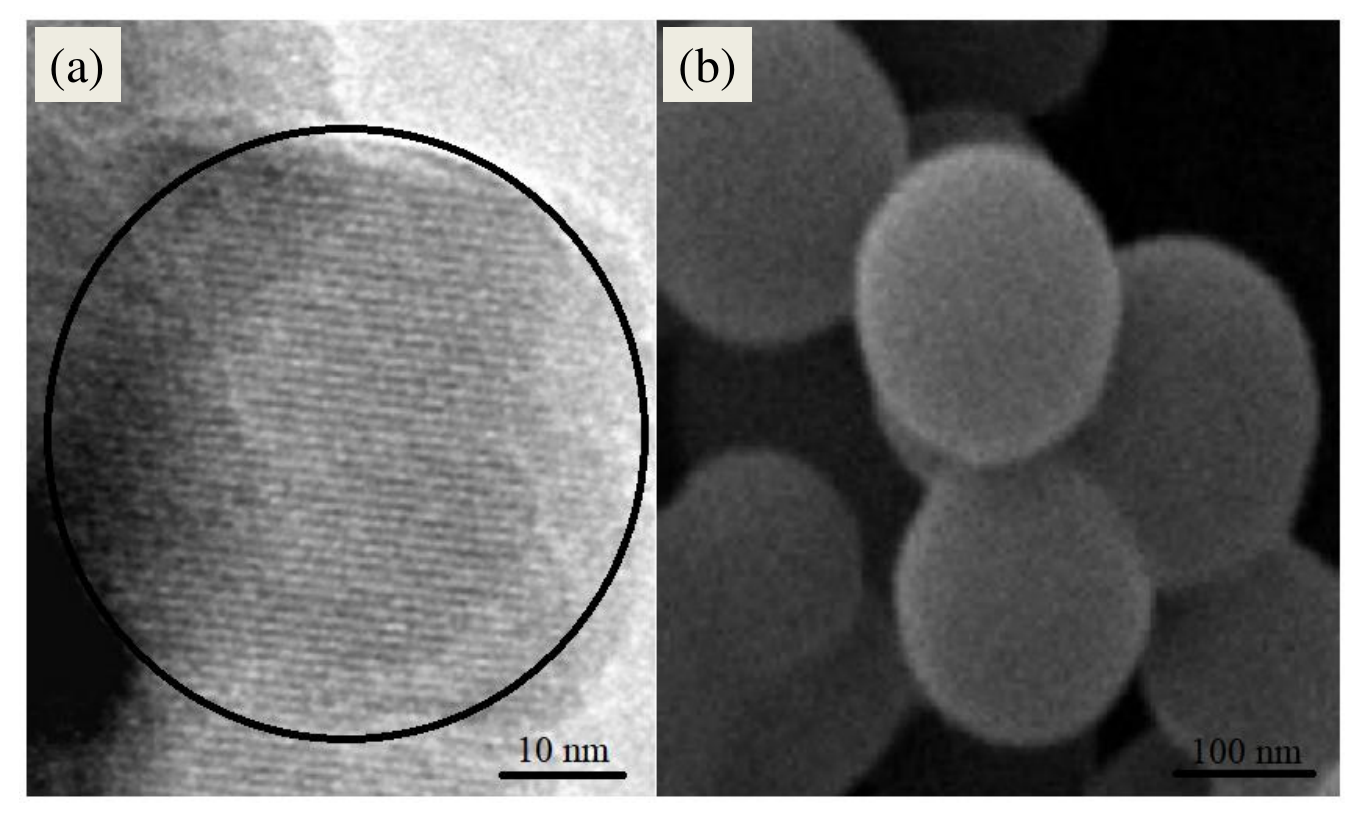}
		\caption{\label{arttype}(Color online)(a) Shows a micrograph of a crystalline Fe$_{3}$O$_{4}$ nanoparticle obtained using a Transmission Electron Microscope (TEM) model 300 kV FEG Tecnai. (b) Shows agglomerates from nanostructures of Fe$_{3}$O$_{4}$/PANI obtained using a Scanning Electron Microscopy (SEM) model 200 kV FEG Quanta. The Fe$_{3}$O$_{4}$/PANI nanoparticles were synthesized under UV radiation with a wavelength of 365 nm at a temperature of 330 K and a times from synthesis of t = 60 minutes.}
		\label{prata}
	\end{center}
\end{figure}

\subsection{Thermal analysis}

To understand the transitions that the nanostructures underwent when they were under UV radiation and to estimate the amount of polymer in a sample, we performed thermogravimetric analysis (TGA) and differential scanning calorimetry (DSC) using a TGA and DSC equipment model LabSys Evo from SETARAM. We carried out the thermograms under an Air atmosphere, with a temperature range between 300 K and 770 K, with a heating rate of 10 K/min. This temperature range was sufficient for us to know when PANI degradation began.

\subsubsection{Thermogravimetric analysis (TGA)}

\textbf{Figs. 4 (a)}, \textbf{(b)}, and \textbf{(c)} show the thermograms for the Fe$_{3}$O$_{4}$/PANI nanostructures synthesized under UV radiation and at temperatures of 310 K, 330 K, and 350 K, respectively. The time of synthesis was varied from t = 0 min. to t = 180 min. in 30 minutes intervals.
\begin{figure}[h]
	\vspace{0.1mm} \hspace{0.1mm}
	\begin{center}
		\includegraphics[scale=0.42]{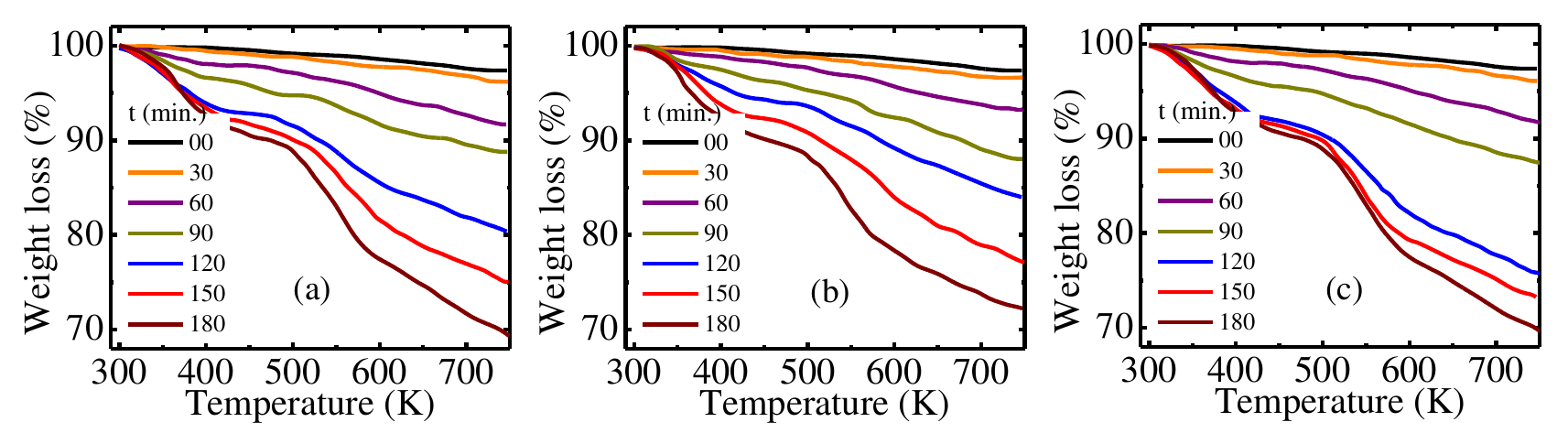}
		\caption{\label{arttype}(Color online)Thermograms for the Fe$_{3}$O$_{4}$/PANI nanostructure samples synthesized under UV radiation and temperatures of 310 K (a), 330 K (b), and 350 K (c). The time of synthesis was varied from t = 0 min. to t = 180 min. in 30 minutes intervals.}
		\label{mala}
	\end{center}
\end{figure}
All thermograms exhibit similar behavior in terms of weight loss. The initial weight loss, around 323 K to 373 K, is associated with the loss of residual water in the material through evaporation [2, 3, 4]. The second loss develops around 373 K to 573 K due to the loss of volatile compounds linked mainly to the PANI chain [3, 4]. The weight loss between 573 K and 770 K is associated with PANI degradation [2, 3, 4]. The amounts of polymer covering the magnetic material increase with increasing times from synthesis.

\subsubsection{Differential scanning calorimetry (DSC)}

\textbf{Figs. 5 (a)}, \textbf{(b)}, and \textbf{(c)} show the calorimetry for the Fe$_{3}$O$_{4}$/PANI nanostructures synthesized under UV radiation and at temperatures of 310 K, 330 K, and 350 K, respectively. The time of synthesis was varied from t = 0 min. to t = 180 min. in 30 minutes intervals.
\begin{figure}[h]
	\vspace{0.1mm} \hspace{0.1mm}
	\begin{center}
		\includegraphics[scale=0.42]{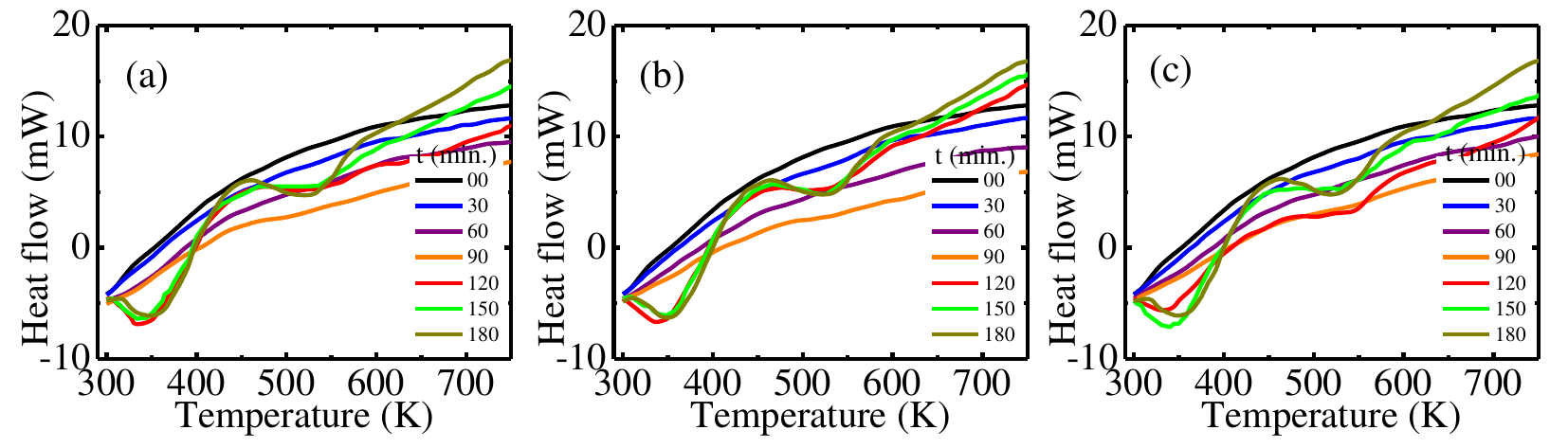}
		\caption{\label{arttype}(Color online) Calorimetry for the Fe$_{3}$O$_{4}$/PANI nanostructures synthesized under UV radiation and at temperatures of 310 K (a), 330 K (b), and 350 K (c). The time of synthesis was varied from t = 0 min. to t = 180 min. in 30 minutes intervals.}
		\label{tira}
	\end{center}
\end{figure}

\textbf{Figs. 5 (a)}, \textbf{(b)}, and \textbf{(c)} show that the nanostructures with time less than t = 90 min. of synthesis do not present endothermic transformations. The results show two endothermic peaks in the time interval between 90 and 180 minutes. The first endothermic peak we attributed to the loss of water that is physically adsorbed on the surface of the materials [2, 3, 4]. Comparing the temperature ranges with the TGA analysis for the same samples, it is possible to observe that the second peak refers to the degradation of polyaniline [2, 3, 4]. 

\subsection{Hysteresis curves}

We performed magnetic measurements using the Vibrating Sample Magnetometry (VSM) method, one of the systems used to study magnetic properties [23]. We used a MicroSense EV7 model VSM adapted to increase its magnetic detection sensitivity, which is in order from 10$^{-8}$ emu. This system is composed of the following subsystems: a pair of electromagnets with a sensitive and stable power supply to produce magnetic fields (H), an electromechanical transducer device with an associated electronic circuit, pickup coils, electronic signal recovery devices, and amplifiers lock-in. We present hysteresis loop measurements performed at room temperature for the Fe3O4/PANI nanostructures synthesized under UV radiation and temperatures of 310 K, 330 K, and 350 K with times from synthesis that varied from t = 0 min. to t = 180 min. in 30 minutes time intervals. In \textbf{Figs. 6 (a)}, \textbf{(b)}, and \textbf{(c)}, we present the results of the hysteresis loops for temperatures of 310 K, 330 K, and 350 K, respectively, which are of high quality [24, 25].
\begin{figure}[h]
	\vspace{0.1mm} \hspace{0.1mm}
	\begin{center}
		\includegraphics[scale=0.42]{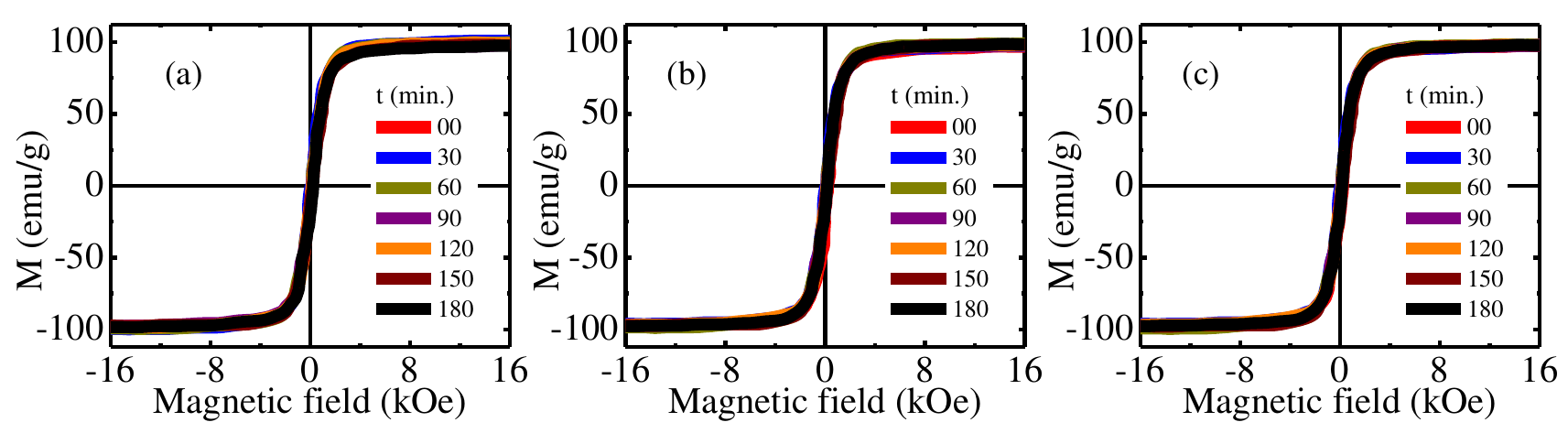}
		\caption{\label{arttype}(Color online) Hysteresis loops for Fe$_{3}$O$_{4}$/PANI nanostructure samples synthesized under UV radiation and at temperatures of 310 K (a), 330 K (b), and 350 K (c). The time of synthesis was varied from t = 0 min. to t = 180 min. in 30 minutes intervals.}
		\label{tira}
	\end{center}
\end{figure}

\section{Conclusion}

The physical entities of exceptional interest in any area are the properties, as they are responsible for different effects that may arise from them. In this work, we report a systematic study on Fe$_{3}$O$_{4}$/PANI nanostructures synthesized under ultraviolet radiation. We highlight the high stability of the magnetic properties of our samples. Furthermore, we present unequivocal evidence that our Fe$_{3}$O$_{4}$/PANI nanostructures synthesized at different temperatures stand out in terms of quality.

\section*{Acknowledgements}
This research was supported by Conselho Nacional de Desenvolvimento Científico e Tecnológico (CNPq), Coordenação de Aperfeiçoamento de Pessoal de Nível Superior (CAPES), Financiadora de Estudos e Projetos (FINEP), and Fundação de Amparo à Ciência e Tecnologia do Estado de Pernambuco (FACEPE). 

\section*{Contributions}
L. H. and T. F. analyzed all the experimental measures, and F. C. and J. H. discussed, wrote and supervised the work.

\section*{Conflicts of interest}
The authors declare that they have no known competing financial interests or personal relationships that could have appeared to influence the work reported in this paper.

\section*{Data availability statement}

Data will be made available on request.

\bibliographystyle{MiKTeX}

\end{document}